\documentclass[aps,prl,twocolumn,unsortedaddress,superscriptaddress,showpacs,10pt]{revtex4-1}

\usepackage{graphicx}
\usepackage[normalem]{ulem}
\usepackage{color,bm}
\usepackage{amsmath,amssymb}
\usepackage[T1]{fontenc} 
\usepackage{siunitx}
\usepackage{url}

\DeclareSIUnit[number-unit-product = {\,}]
\cal{cal}

\newcommand{\beq}{\begin{equation}}
\newcommand{\eeq}{\end{equation}}

\usepackage[version=3]{mhchem}

\begin{document}
\title{Energy dependent stereodynamics of the Ne($^3$P$_2$)+Ar reaction}
\author{Sean D. S. Gordon}
\author{Junwen Zou}
\author{Silvia Tanteri}
\affiliation{Institute for Chemical Sciences and Engineering, Ecole Polytechnique F\'ed\'erale de Lausanne (EPFL), 1015 Lausanne, Switzerland}
\author{Justin Jankunas}
\affiliation{Institute for Chemical Sciences and Engineering, Ecole Polytechnique F\'ed\'erale de Lausanne (EPFL), 1015 Lausanne, Switzerland}
\affiliation{deceased}
\author{Andreas Osterwalder}
\email{andreas.osterwalder@epfl.ch}
\affiliation{Institute for Chemical Sciences and Engineering, Ecole Polytechnique F\'ed\'erale de Lausanne (EPFL), 1015 Lausanne, Switzerland}

\date{\today}

\begin{abstract}
The stereodynamics of the Ne($^3$P$_2$)+Ar Penning and Associative ionization reactions have been studied using a crossed molecular beam apparatus. The experiment uses a curved magnetic hexapole to polarise the Ne($^3$P$_2$) which is then oriented with a shaped magnetic field in the region where it intersects with a beam of Ar($^1$S). The ratios of Penning to associative ionization were recorded over a range of collision energies from \SI{320}{\per\centi\metre} to \SI{500}{\per\centi\metre} and the data was used to obtain $\Omega$ state dependent reactivities for the two reaction channels. These reactivities were found to compare favourably to those predicted in the theoretical work of Brumer \emph{et al.}
\end{abstract}

\maketitle
\section{Introduction}
Collision processes of electronically excited atoms have long been used as rich systems to study elementary energy transfer processes.
Two basic reaction channels are available for energy disposal if the internal energy of the excited atom A* is higher than the ionization energy of another atom, B:
\begin{eqnarray}
\text{A}^*+\text{B}&\longrightarrow& \text{A}+\text{B}^++e^-\label{eqn:reactions} \\
&\longrightarrow& \text{AB}^++e^-.\label{eqn:reactions2}
\end{eqnarray}
The first of these reaction is called Penning ionization (PI), the second is associative ionization (AI), also known as the Hornbeck-Molnar process.\cite{hornbeck1951mass}

A plethora of data is available for the PI channel.\cite{miller1970theory,siska1993molecular,ohoyama1999evidence,kohmoto1977penning,hotop1981penning,niehaus1973penning} 
Quantities of physical relevance such as the state specific total ionization cross sections and electron angular distributions have been obtained, yielding further insight into the energy transfer process.\cite{parr1982velocity,driessen1990pure,tang1972velocity,hotop1971angular,hoffmann1979interpretation,hotop1974analyses} 
Later, it was found that PI experiments are ideally suited for use in very low collision energy experiments. 
The ionization process is barrierless and the cross sections display resonances at low collision energy.\cite{henson2012observation}
These cold collision experiments have become a proving ground for ultracold collision dynamics as the metastable atoms can be manipulated and controlled by external magnetic fields.\cite{leo2001ultracold,jankunas2014dynamics,osterwalder2015merged,henson2012observation,%
lavert2014observation}

The branching ratio between PI and AI has been investigated for a number of different collision systems, and it is found for most that the PI process is favoured.\cite{neynaber1976penning,niehaus1973penning,aguilar1985velocity} 
The PI channel is thought to dominate in the collisions of rare gases due to the very weak bond formed between bound state rare gas ions which typically only supports a small number of bound rovibrational states.\cite{siska1993molecular} 
One compelling question that comes out of the competition between reactions~\ref{eqn:reactions} and \ref{eqn:reactions2} is how the angular momentum polarization of the excited electron affects the reaction outcome.  
Both PI and AI involve a form of coupling between electronic states of the impinging A* and B, but the exact dependence of the complexation mechanisms on angular momentum have received scant attention in the literature.\cite{ohoyama2008steric}  

Details of the collision process itself can be assessed by controlling the spatial orientation of reactants and investigating the properties of the reaction products.
The study of stereodynamics can be experimentally realised by orienting at least one of the reactants with inhomogeneous electric, magnetic or optical fields (See Ref.~\cite{aoiz2015new} and citations therein).
For example, by orienting a polarised reactant in a magnetic field one sets the direction of the magnetic moment $\vec{\mu}$.
Since the relative spatial positions of $\vec{\mu}$ and the angular momentum $\vec{J}$ vector are well known, polarising the magnetic moment allows the keen experimentalist to manipulate the relative populations of states with different projections of $\vec{J}$, labeled $\Omega$, on the inter-atomic axis.

In the present experiment the branching ratio of AI vs. PI in collisions of metastable Ne($^3P_2$) (henceforth designated as Ne*) with ground state Ar in a controlled magnetic field is investigated.
This reaction, and specifically the energy dependent branching ratio between AI and PI has been at the focus of previous theoretical investigations concerning the question of coherent control of the reaction.\cite{arango2006cold,bell1968penning} 
An important result from these papers are the $\Omega$-specific cross sections for AI and PI.
These results predict that the reactivities $R_\Omega=\sigma^{AI}_\Omega/\sigma_\Omega^{PI}$ are small and of similar magnitude for $\Omega=$0 and $\Omega=1$ but larger for $\Omega=2$.
No experimental evidence produced until now could support this claim.

We here use a tuneable magnetic field in the reaction zone to orient the polarised Ne* exiting a curved hexapole magnetic guide.\cite{woestenenk2001construction,jankunas2016oriented} 
The transmission probability for the Ne* atoms in the guide depends on the magnetic quantum number $m_J$ (see supp. matt. and references \onlinecite{osterwalder2015merged,Jankunas:2015cka} and \onlinecite{vandeMeerakker:2012ft}). 
After exiting the guide, the Ne* $\vec{J}$ adiabatically orients to the tuneable magnetic field in the scattering chamber. 
The oriented Ne* beam is then crossed at right angles with a beam of ground state Ar atoms leading to PI and AI of the Ar atom. 
The relative intensities of the two resulting signals as a function of the magnetic field angle give access to the $\Omega$ state dependent reactivities. 
Product ions are detected in a mass spectrometer and recorded as a function of orientation angle and of collision energy. 

The polarization of the Ne* sample is described using a density matrix formalism giving the population of states ordered by the quantum number $m_J$,  the projection of $\vec{J}$ on the magnetic field axis.\cite{blum2012density}
For the collision process itself, the relevant populations are those for states labelled by $\Omega$.
The calculation of the populations $p_\Omega$ is obtained through a frame transformation from the laboratory frame of reference into the molecular frame of reference, using the externally defined direction of the magnetic field and the relative velocity of Ar and Ne*.
Because the $p_{m_J}$ are fixed by the guiding dynamics in the curved magnetic hexapole, the rotation of the magnetic field at constant relative velocity leads to different proportions of $p_\Omega$.
By measuring relative cross sections of PI and AI as a function of the field direction and the collision energy we make use of our complete control over all degrees of freedom in the reaction and obtain a two-dimensional reactivity map which is fitted to energy- and $\Omega$-dependent reactivities $R_\text{AI/PI}(\theta,E)$.

\section{Experimental }
The apparatus used here has been described in detail elsewhere in the context of merged-beam experiments.\cite{jankunas2014dynamics,osterwalder2015merged,Bertsche:2014kl,Jankunas:2015fm}
There are, however, a few notable differences. Most importantly, the setup is operated in crossed-beam configuration.

A velocity-tuned supersonic expansion of Ne($^3P_2$) atoms is generated by a dielectric barrier discharge from a temperature controlled Even-Lavie valve\cite{luria2011generation,luria2009dielectric} (repetition rate \SI{30}{\hertz}; stagnation pressure \SI{8}{\bar}). 
Metastable Ne atoms emerging from the discharge are skimmed and enter a \SI{1.8}{\meter} long curved magnetic guide.
Only the $^3$P$_2$ state is sufficiently long-lived and efficiently guided in the magnetic field and reaches the interaction zone\nocite{Brooks:1976kda,Anonymous:1kzDvH3B,Brooks:1966jc,Bulthuis:1995bb,Nichols:2015hl}.\cite{suppmattcite,watanabe2006characterization,osterwalder2015merged}
The valve temperature is controlled through cooling with liquid nitrogen and heating with a coil on the valve body.
Mean beam velocities between \SI{500}{\metre\per\second} and \SI{800}{\metre\per\second} are obtained. 
\begin{figure}
\centering
\includegraphics[width=0.48\textwidth]{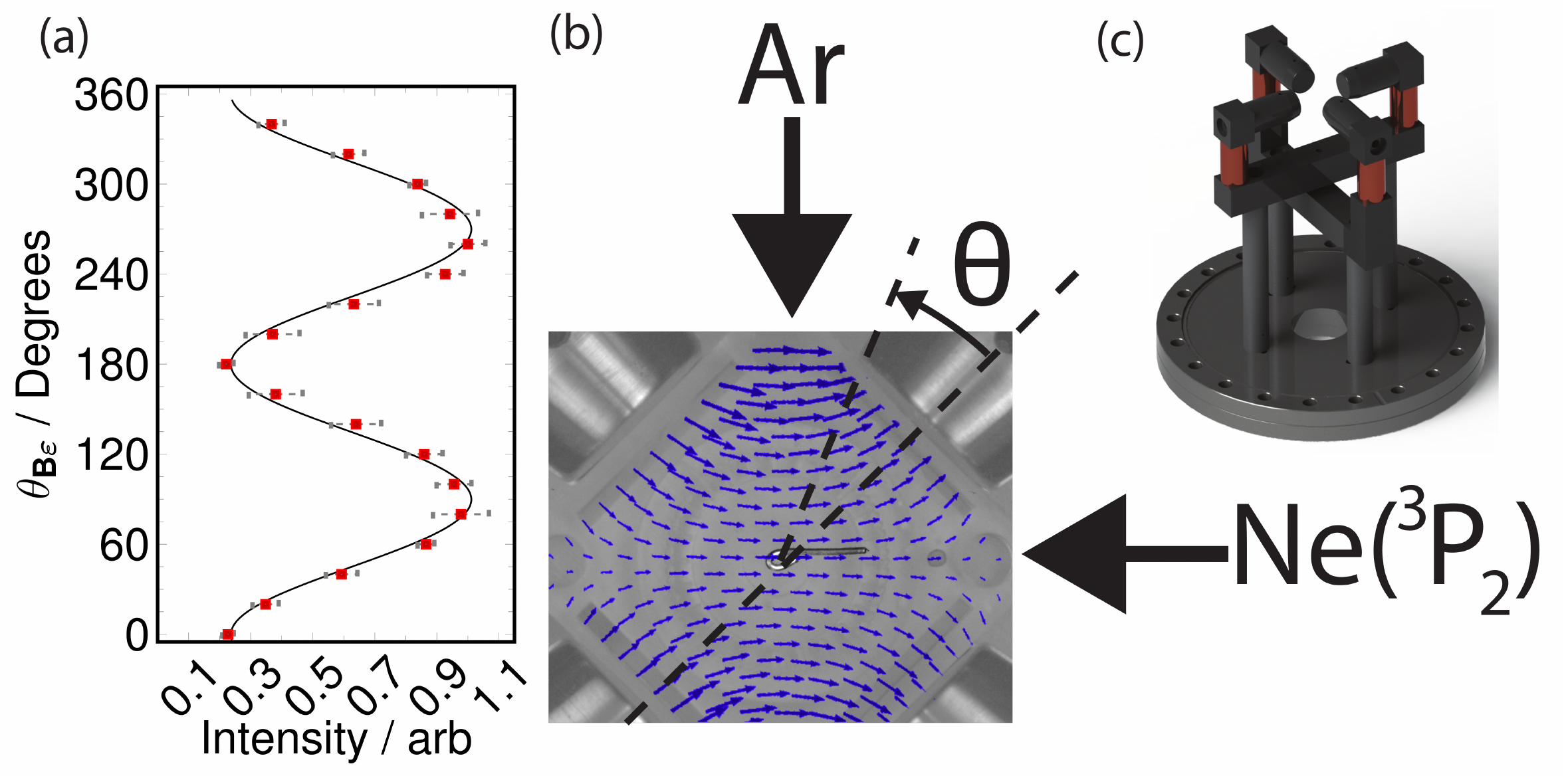}
\caption{\label{fig1} a) 1+2 REMPI signal of Ne* as a function of the angle between laser polarization and magnetic field. b) Schematic of the experimental arrangement. A beam of Ne($^3P_2$) enters the reaction zone from the left and is intersected by a beam of Ar. The photograph shows the ends of the electromagnets and a calculated field map for a magnetic field direction of $\theta=135^\circ$, where $\theta=0$ is defined along one of the magnets as shown. Product ions are extracted perpendicularly to the plane of the Figure. c) Rendering of the solenoid magnets used for the orientation field. }
\end{figure}

The Ne* beam crosses a skimmed, pulsed, supersonic expansion of Ar atoms, produced from a room-temperature general valve operated with a backing pressure of \SI{2.5}{\bar}, at ninety degrees in the center of the magnet assembly shown in Fig. \ref{fig1}.
Panel \ref{fig1}c shows the arrangement of the two solenoid magnets, oriented at ninety degrees with respect to each other and at 45 degrees to both beams, as shown in panel \ref{fig1}b.b
The direction of the field is controlled by applying different currents to each of the solenoids in a manner that the amplitude at the center is always $\approx$ 10 mT.
For example, an angle of $\theta=0^\circ$ is achieved by applying current only to the magnets along the $\theta=0^\circ$ axis while equal current on both magnets produces a field oriented at $\theta=45^\circ$.
Measurements of the degree of control using a CCD camera and compass needle, as shown in Fig. \ref{fig1}b, have shown that the field direction can be repeatedly set with a precision better than \SI{5}{\degree}. 

Collision products, Ar$^+$ and NeAr$^+$ are accelerated in a ToF mass spectrometer and recorded separately on an MCP detector.
Signals from several thousand beam pulses are accumulated at each orientation angle and Ne* velocity. 

\section{Results and Discussion}
\begin{figure}
\centering
\includegraphics[width=0.9\columnwidth]{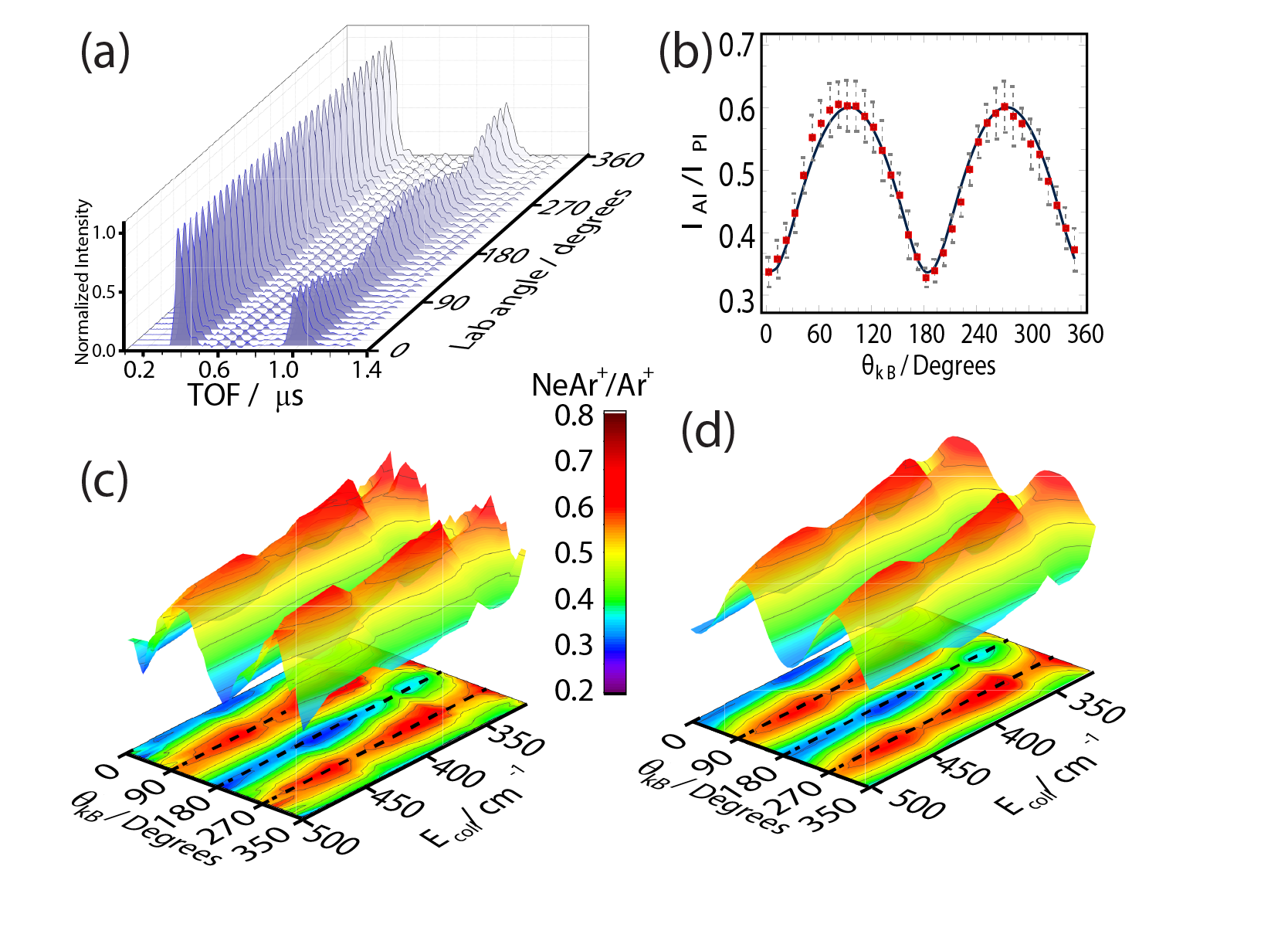}
\caption{\label{fig2} a): Time-of-flight spectra at a collision energy of 390 cm$^{-1}$ as a function of the magnetic field-direction in the laboratory reference frame. All traces are normalized to the Ar$^+$ peak at 400 ns. b) Ratio between the AI and PI signal at a collision energy of 390 cm$^{-1}$ as a function of $\theta_{kB}$. c) and d) Experimental (c) and fitted (d) reactivity as a function of $\theta_{kB}$ and collision energy. } 
\end{figure}

Fig. \ref{fig2}a shows normalised time-of-flight spectra for the Ne*+Ar reaction recorded at a collision energy of \SI{390}{\per\centi\metre}, as a function of the magnetic field direction.
Each trace in Fig. \ref{fig2}a contains only the early Ar$^+$ peak and the later NeAr$^+$ peak.
The angle-dependent ratio between these two peaks is plotted in Fig. \ref{fig2}b, which thus shows $\frac{\sigma_{tot}^{AI}}{\sigma_{tot}^{PI}}=\frac{\ce{NeAr+}}{\ce{Ar+}}$ for this particular collision energy.
Joining curves such as the one in panel \ref{fig2}b for the collision energy range 320-500 cm$^{-1}$ produces the surface shown in panel \ref{fig2}c.

 The PI channel is maximised when the field direction is parallel to the relative velocity and AI is maximised when the field is perpendicular to the relative velocity. 
The signal intensity for either the AI or PI reaction channel is the sum of the state dependent reactivities weighted by the populations,
\begin{equation}\label{eqn:fit}
I_\text{AI;PI}(\theta_{\bm{k}\text{\textbf{B}}})\propto\sum\limits_{\Omega=0}^{J}p_\Omega(\theta_{\bm{k}\text{\textbf{B}}})\sigma_\text{AI;PI}^\Omega.
\end{equation}
$\theta_{\bm{k}\text{\textbf{B}}}$ is the angle between the interatomic velocity $\vec{k}$ and the magnetic field axis, $\sigma_\text{AI;PI}^\Omega$ are the $\Omega$ state dependent cross sections of the PI or AI channel and $p_\Omega(\theta_{\bm{k}\text{\textbf{B}}})$ the population of the $\Omega$ quantum number. 
The $p_\Omega(\theta_{\bm{k}\text{\textbf{B}}})$ populations are not measured directly but are derived from the $p_{m_J}$ that are defined in the laboratory frame and can be determined spectroscopically.
These are then transformed by a rotation of the reference frame from the laboratory into the molecular frame, using the Wigner d-matrices as detailed in the supplementary material.\cite{zare2013angular}
The $p_{m_J}$ were obtained by recording the 1+2 REMPI signal of Ne($^3P_2$) at $\approx$\SI{320}{\nano\metre} using linearly and circularly polarised light, extracting polarization moments, and transforming into state populations.\cite{Rakitzis:2010ft,zare2013angular,jankunas2016oriented,rakitzis1999photofragment} 
A plot of the experimental and fitted REMPI signal intensity for linearly polarised light as a function of the laser polarization angle is shown in Fig. \ref{fig1}a. 
Here, the direction of the magnetic field was kept constant while the laser polarization was rotated.
An identical result was obtained when keeping the laser polarization fixed but rotating the magnetic field, demonstrating that the reorientation of the Ne* atoms in the magnetic field indeed is an adiabatic process that retains the $p_{m_J}$ populations.
The extracted state populations from the REMPI measurements were determined to be \sisetup{separate-uncertainty} $p_0=\num{0.087(15)}$, $p_1=\num{0.75(2)}$ and  $p_2=\num{0.163(25)}$. 
These values were found to change inappreciably over the Ne* velocity range.
The experimental ratio from Fig. \ref{fig2}c can be used in conjunction with the above theoretical framework to obtain relationships between the state dependent reactivities, $R_\text{AI/PI}^\Omega$. 

\begin{figure}
\centering
\includegraphics[width=0.45\textwidth]{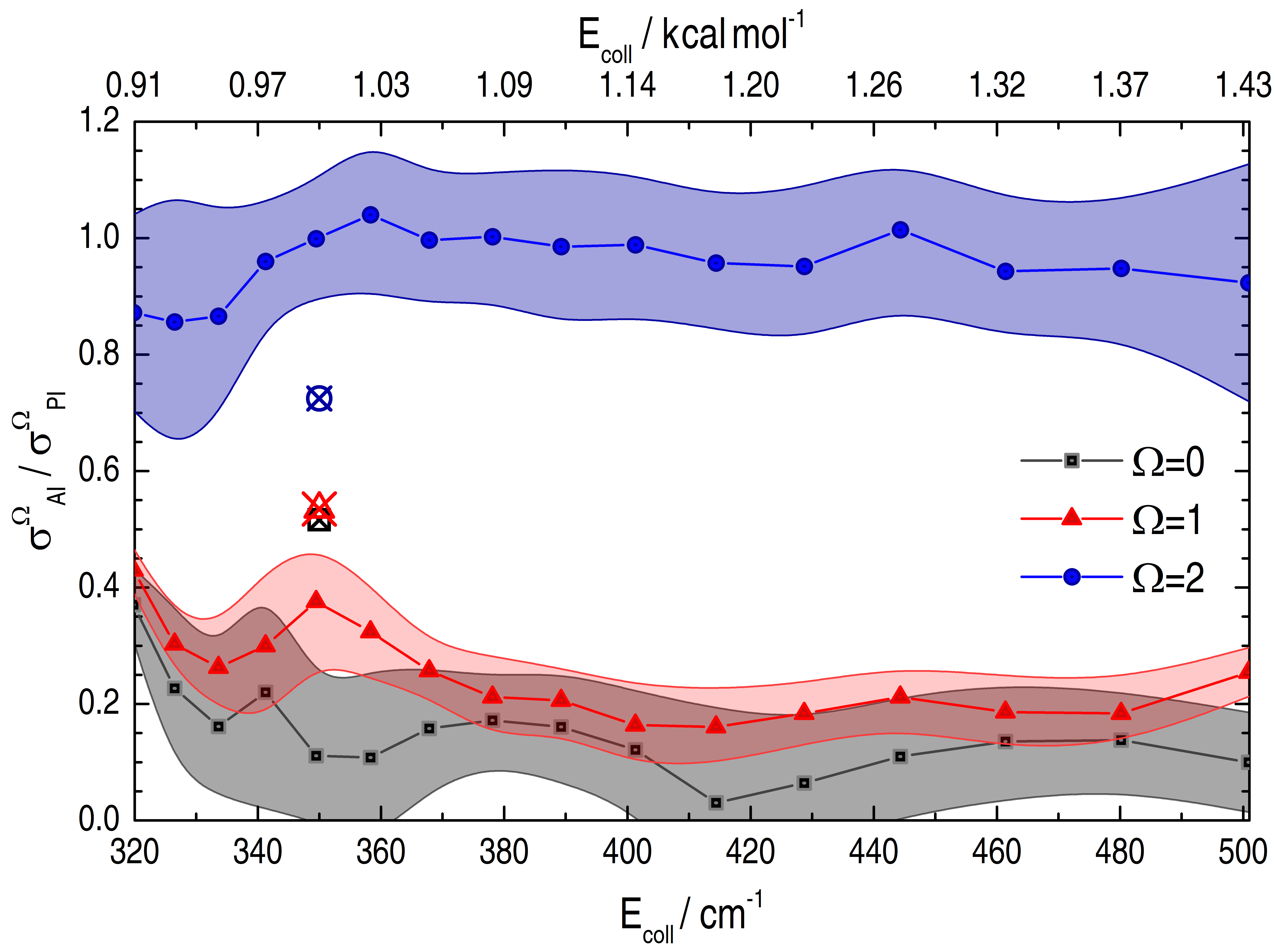}
\caption{\label{fig3} Experimental, energy dependent reactivities for $\Omega=0$ (black square), $\Omega=1$ (red triangle), and $\Omega=2$ (blue circle). Confidence intervals enclose the sample standard deviation. Theoretical results for a collision energy of 350 cm$^{-1}$ (1\,kcal\,mol$^{-1}$) are shown for reference (crossed symbols).\cite{arango2006coherent,arango2006cold}   }
\end{figure}

It is worth pointing out that the present measurements of the AI/PI ratio do not yield absolute cross sections for the PI and AI channels.
This rules out the possibility of directly solving equation~\ref{eqn:fit} and prevents determination of the absolute values of $\sigma_\text{AI;PI}^\Omega$. 
Instead, we use the experimental reactivities $R_\text{AI/PI}(\theta,E)$ to extract the relevant $\Omega$ dependent reactivities.
The total reactivity is given as
\begin{eqnarray}\label{eq:react}
R_\text{AI/PI}(\theta,E)&=&\frac{I_\text{AI}(\theta_{\bm{k}\text{\textbf{B}}})}{I_\text{PI}(\theta_{\bm{k}\text{\textbf{B}}})} =\frac{\sum\limits_{\Omega=0}^{J}p_\Omega(\theta_{\bm{k}\text{\textbf{B}}})\sigma_\text{AI}^\Omega}{\sum\limits_{\Omega=0}^{J}p_\Omega(\theta_{\bm{k}\text{\textbf{B}}})\sigma_\text{PI}^\Omega},
\end{eqnarray}
and this is plotted, using the fitted values, in Fig. \ref{fig2}d to reproduce the experimental plot in panel \ref{fig2}c.
The fit was obtained using a non-standard Monte-Carlo fitting algorithm, described in detail in the supplementary material. 
Briefly, the target function equation~\ref{eq:react} for different angles and collision energies was fitted to the experimental results using the experimental populations $p_{m_J}$, appropriate frame rotation, and adjusting the six cross sections $\sigma_\text{AI;PI}^\Omega$.
This procedure provides the desired $\Omega$-specific reactivities $\sigma_\text{AI}^\Omega/\sigma_\text{PI}^\Omega$.
As detailed in the supplementary material, a direct fit of these parameters is rendered difficult by the high degree of correlation between them.

To illustrate the quality of the fit we show the best result for the specific collision energy of 390 cm$^{-1}$ in Fig. \ref{fig2}b. 
Energy and $\Omega$ dependent reactivities are shown in Fig.~\ref{fig3}. 
The results for $\Omega$=0,1, and 2 are shown as black, red, and blue traces, respectively.
In each case the shaded area shows the confidence interval for the fit (one standard deviation; see supplementary material for details).
Results from previous theoretical calculations of $R_\Omega$ are shown for reference in Fig. \ref{fig3} for a collision energy of \SI{1}{\kilo\cal\per\mole}\cite{arango2006coherent,arango2006cold} 
Satisfactory agreement is observed in particular with regards to the similar reactivities $R_0$ and $R_1$ in comparison with $R_2$.

\begin{figure}
\centering
\includegraphics[width=0.4\textwidth]{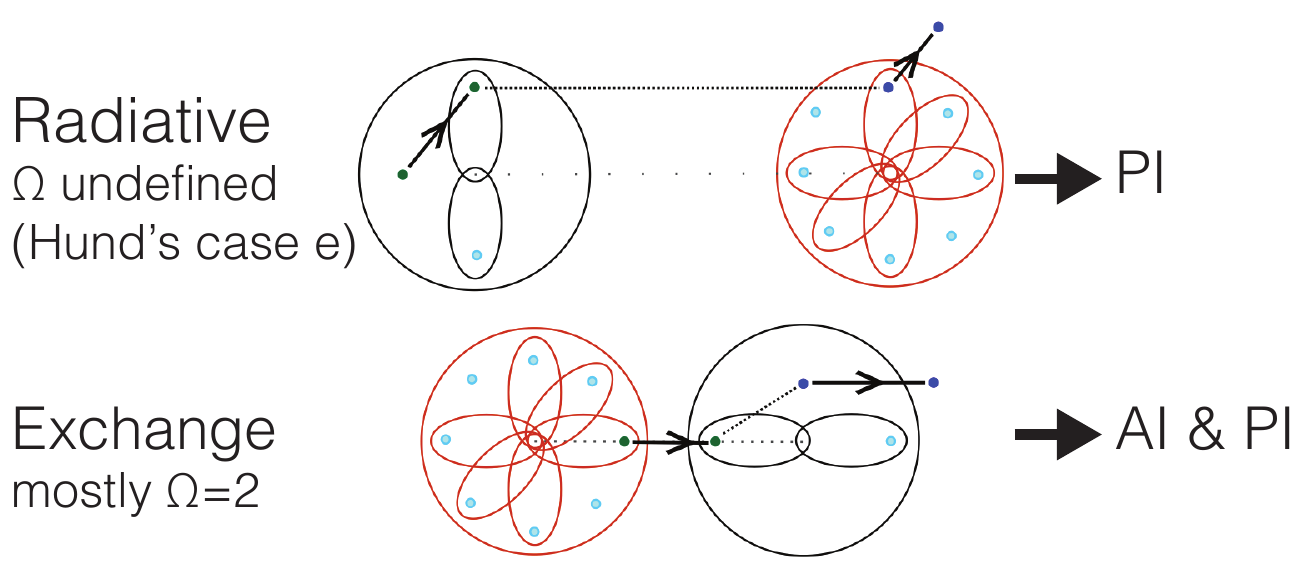}
\caption{\label{fig4} Cartoons for the postulated mechanisms: radiative mechanism (top), operates at long range and leads to PI; exchange mechanism (bottom) operates at close range and leads to AI and PI. At long range $\Omega$ is not a good quantum number and the radiative mechanism has similar cross sections for all $\Omega$ states; higher cross sections for $\Omega=2$ are predicted for the exchange mechanism.}
\end{figure}

Two independent reaction mechanisms have been proposed which can result in either PI or AI, these are shown pictorially in Fig. \ref{fig4}.
The $\Omega$ dependent reactivities shed light on the relative importance of each mechanism for forming either the AI or PI products. 
The two competing pathways, with a different orientation dependence, are referred to as the radiative and the exchange mechanisms.\cite{siska1993molecular,aguilar1985velocity} 
The exchange mechanism (bottom illustration in Fig. \ref{fig4}), believed to be dominant in the Ne*+Ar collision, involves the outer $p$ electron of Ar being transferred to the vacant $p$ orbital of Ne* and causing the ejection of the Penning electron.\cite{miller1977unified,hotop1969reactions} 
This mechanism requires orbital overlap between the partially occupied Ne* $p$ orbital and a $p$ orbital of the Ar. Thus, it is important particularly at short interatomic distances. 
Favourable orientation of the singly occupied $p$-orbital in Ne* with respct to the Ar makes this process more efficient for $\Omega$=2 projections than for $\Omega$=0 and 1.


In the radiative mechanism (upper illustration in Fig. \ref{fig4}) the Ne* $s$ electron transfers to its $p$ orbital which leads to ionization of Ar through a long range coupling. \cite{aguilar1985velocity,bussert1985ionizing}
The radiative process can also occur at long range where the system is best described as a Hund's case (e) and $\Omega$ is not a good quantum number, the orientation of the orbital in this mechanism is therefore expected to be unimportant.\cite{aguilar1985velocity}
While the exchange mechanism proceeds through a complex and thus can lead to AI, the radiative mechanism results exclusively in PI. The relative reactivities therefore contain information about the importance of the long and short range mechanisms on the ion yield of PI and AI.

We find experimentally that for $\Omega=2$ ($p$ orbital pointing towards Ar atom), AI has approximately equal cross section to that of PI across the energy range ($\sigma^{\Omega=2}_\text{AI}/\sigma^{\Omega=2}_\text{PI}\approx1$). 
In this configuration, as shown in the bottom panel of Fig.~\ref{fig4}, the Ne* and Ar atoms can undergo complexation through the exchange mechanism. 
When the $p$ orbital lies towards the incoming Ar, the configuration favours exchange and the internal energy transfer from the excited electron to the bound rovibrational states in the complex decides the cross section for the AI and PI channel. 
When $\Omega=0$ or 1, the argon $p$ orbital overlaps less well with that of the neon $p$ orbital which suppresses complex formation.\cite{aguilar1985velocity,driessen1990pure} 
At long range where $\Omega$ is not a good quantum number, the radiative mechanism, leading to PI, is favoured. 
In this case there is little chance of forming a bound state complex as the ionization occurs before complexation can occur.\cite{aquilanti1986hyperspherical,aguilar1985velocity} Justified by previous experimental results, calculations have often neglected the radiative contribution.\cite{miller1977unified,bussert1985ionizing} However, our data suggests that this approximation may not be valid under the present circumstances.

At the lowest collision energies sampled here, the AI channel is increasing in importance for $\Omega=0$ and 1. 
Since these quantum numbers mean that the singly occupied $p$ orbital is not pointing toward the atom, this trend may mean either that at these energies the long range radiative mechanism is becoming less favoured, or that at low collision energies the propensity for complexation increases even though the orbital overlap is less ideal. 
 At higher collision energy, around \SI{260}{\milli\electronvolt}, the AI channel disappears completely.\cite{arango2006coherent} 
 There, the energy of the collision is so high that the complex can not form any bound states, so even if the system proceeds through the exchange mechanism and complexation, the only possible outcome is PI whose cross section is predicted to steadily increase at higher collision energy. 
 Calculations predict that at ultracold temperatures the reaction is entirely dominated by AI, meaning that the radiative process should lose it's importance entirely.\cite{arango2006cold}
 Stereoydynamics studies in a merged beams configuration are now planned to assess this and obtain reactivities in the cold regime with a collision energy < \SI{1}{\kelvin}.

\section{Conclusions}
The ratio of Penning to associative ionization for the reaction of Ne*+Ar across the energy range \SI{320}{\per\centi\metre} to \SI{500}{\per\centi\metre} has been measured using a crossed-beam apparatus. The experimental results provide ratios of state dependent cross sections which have been shown to agree favourably with theory. The Penning process is preferred across the energy range and becomes ever more dominant as the collision energy increases, the probability of forming a complex below the dissociation threshold diminishes as the collision energy is increased as the ionised complex can not support a large number of rovibrational levels. The state dependent cross sections do not show large variation across the collision energy range. At the lowest collision energies sampled the importance of the $\Omega=2$ channel diminishes as the proportion of AI in the total ionization cross section increases. The importance of the $\Omega=2$ channel for AI agrees favourably with the proposed exchange mechanism in which overlap of the Ne* $p$ orbital with the Ar leads to complex formation and hence AI.

\begin{acknowledgments}
This work is funded by EPFL and the Swiss Science Foundation (project number 200021\_165975).
\end{acknowledgments}


%

\end{document}